\documentclass[pra,twocolumn,superscriptaddress,showpacs]{revtex4}
\usepackage{bbm}
\usepackage{amsmath, amssymb}
\usepackage{color}
\usepackage{graphicx,epsfig}
\usepackage{pifont}
\usepackage{subfigure}
\usepackage{amsmath, amssymb, graphics}

\def\ket#1{\lvert\nobreak#1\nobreak\rangle}
\def\bra#1{\langle\nobreak#1\nobreak\rvert}
\def\av#1{\langle\nobreak#1\nobreak\rangle}

\begin{document}

\title{Generating strong squeezing in the dispersive regime of the quantum Rabi model}
\author{Chaitanya Joshi}
\email{chaitanya.joshi@york.ac.uk}
\affiliation{Department of Physics and York Centre for Quantum Technologies, University of York, Heslington, York, YO10 5DD, UK }
\author{Elinor K. Irish}
\affiliation{Physics and Astronomy, University of Southampton, Highfield, Southampton, SO17 1BJ, UK}
\author{Timothy P. Spiller}
\affiliation{Department of Physics and York Centre for Quantum Technologies, University of York, Heslington, York, YO10 5DD, UK }
\date{\today}

\begin{abstract}
We present a protocol to generate a large degree of squeezing of a boson (light) field mode strongly coupled to a two-level system in the dispersive regime. Our protocol exploits the strong dispersive coupling to introduce a time dependent frequency change of the boson field. With an appropriately timed sequence of sudden frequency changes, the quantum noise fluctuations in one quadrature of the field can be reduced well below the standard quantum limit, with a correspondingly increased uncertainty in the orthogonal quadrature. Even in the presence of realistic noise and imperfections, the protocol should be capable of generating substantial squeezing with present experimental capabilities. 
\end{abstract}

\maketitle
\section{Introduction}
Quantum technology promises to provide a robust platform combining otherwise contrasting degrees of freedom to achieve various tasks of quantum information and computation. Light-matter coupled quantum systems have received particular attention in the bid to engineer scalable quantum platforms \cite{zelx13,mhde13,gkur15}. In these hybrid systems a two-level system (qubit) is coupled to a bosonic field. Theoretically, the basic building block for describing such light-matter systems is the quantum Rabi model, which takes the following form under the dipole approximation \cite{iira36},
\begin{equation}\label{fullrabi}
H_{\rm Rabi}=\omega \hat{a}^{\dagger}\hat{a}+\frac{\Omega}{2} \hat{\sigma}_{z} + g(\hat{a}^{\dagger}+\hat{a})(\sigma^{+}+\sigma^{-}).
\end{equation}
Here, $\hat{a}$ and $\hat{a}^{\dagger}$ are the annihilation and creation operators for the bosonic field of frequency $\omega$, $\hat\sigma^{\pm}=(\hat\sigma_{x}\pm i\hat\sigma_{y})/2$ with $\hat\sigma_{x,y,z}$ the Pauli matrices for the two-level system, $\Omega$ is the energy level splitting between the two levels, and $g$ denotes the coupling strength between the bosonic mode and the two-level system (assumed to be positive). 

In conventional cavity QED settings the light-matter coupling strength $g$ is several orders of magnitude smaller than the transition frequencies $\omega,\Omega$. In addition, if the system is near resonance, such that $\omega \sim \Omega$, the full Rabi model can be simplified by applying the rotating wave approximation (RWA). Under this approximation the so-called ``counter-rotating'' terms in Eq.~\eqref{fullrabi} can be neglected, leading to the much simpler and readily solvable Jaynes-Cummings (JC) model~\cite{etja63}
\begin{equation}\label{rabirwa}
H_{\rm JC} = \omega \hat{a}^{\dagger}\hat{a} + \frac{\Omega}{2} \hat{\sigma}_{z} + g(\hat{a}^{\dagger}\sigma^{-} + \hat{a}\sigma^{+}).
\end{equation}
However, the validity of the Rabi model is not only restricted to cavity QED setups. The original Rabi model is a ubiquitous physical model capable of describing a wide variety of other physical systems, including trapped ions \cite{rbla12,jped15}, qubit-coupled nanomechanical resonators \cite{ekir03,Rouxinol2016}, and circuit QED systems \cite{abla04,awal04,Rigetti2012,Barends2013,Braumuller2016,Yan2016}. Circuit QED architectures are particularly interesting because it is possible to reach physical regimes where the light-matter coupling strength $g$ becomes a sizeable fraction of the transition frequency of the boson field and/or the two-level system, meaning that the RWA is no longer a valid approximation. The influence of non-RWA terms has been studied theoretically for some time \cite{pneu96,rfbi01,ekir03,ekir05,gjoh06,jlar07,ekir07,jhau08}. However, with recent experimental advances in reaching ``ultra-strong'' and ``deep strong'' coupling regimes of light and matter \cite{pfor16,fyos16}, investigations of the full Rabi model are attracting increasing attention \cite{tnie10,Rigetti2012,esol11,jcas10,bpero10,dbra11,jqyo11,pnata11,dball12,jped15}. It has been clearly shown that in these strongly coupled quantum systems the simplified JC model no longer applies and it becomes necessary to consider the full Rabi model to capture the relevant physics. The significance of non-RWA terms has also been elucidated in many-body extensions of the full Rabi model, both in the equilibrium \cite{msch12} and non-equilibrium \cite{msch16,cjosh16} settings. 

Another regime of significant practical interest in which the Rabi model can be substantially simplified is the so-called ``dispersive regime'' \cite{abla04}. In the dispersive limit, the qubit and the boson field are far detuned compared to the light-matter coupling strength $g$, {\it i.e.} $g \ll \lvert \Omega-\omega \rvert$. This regime is widely considered in experiments, particularly in circuit QED, as it allows a non-demolition type measurement of the qubit by probing the resonator \cite{awal04}. Although the dispersive approximation is often applied together with the RWA \cite{awal04}, in this work we consider the dispersive limit without making the RWA  \cite{ekir03, dzue09} and show that it can be used to create squeezed states of the field mode. 

The generation of squeezed light has attracted much attention for various applications, including high-precision quantum measurements \cite{cmca81,jaasi13,matay13} and quantum communication \cite{slbr05}. In the quantum optical domain, squeezed light has been more commonly generated using nonlinear optical processes, including degenerate parametric amplification and degenerate four-wave mixing \cite{hpyu78,resl85,rmsh86,la87,walls}. These nonlinear processes require large optical nonlinearities, low intracavity and detection losses, and low phase noise \cite{rlou87,uland,rgarc16}. Over the years, experimental progress has made it possible to generate squeezed states of light in a variety of different experimental setups \cite{rgarc16}, including optical parametric oscillators \cite{teber10}, superconducting cavities \cite{macas08}, and optomechanical cavities \cite{tppu13}. 

In this work we embark on an alternative route to generate squeezed states of a cavity field mode dispersively coupled to a two-level system. It has been shown that in the dispersive regime beyond the RWA, the ground state of the Rabi model exhibits one-mode squeezing of the boson field \cite{dzue09}, but the degree of squeezing is very small. However, it has been known for a long time that any nonadiabatic change in the frequency of a harmonic oscillator (boson field) results in squeezing of the state of the oscillator \cite{cflo90,gsaga91}. The degree of squeezing is particularly pronounced if the frequency change is sudden \cite{jjan86}. Moreover, it is possible to use periodic sudden jumps between two frequencies to produce arbitrarily large squeezing of the field mode \cite{jjan92,vvdo93,amzag08}. We show that this strategy can be used in the context of the dispersive quantum Rabi model to generate significant squeezing of the field mode. Comparing analytical predictions using the dispersive approximation with numerical simulations of the same protocol using the full Rabi Hamiltonian, we find that, remarkably, an even larger degree of squeezing arises in the latter case. An analysis of the effects of noise and imperfections suggests that considerable squeezing could be achieved with existing experimental capabilities in circuit QED.
%
%
%
%
%

The paper is organized as follows. In Sec.\ref{sec:disp} we introduce the dispersive theory describing the interaction between a quantized bosonic mode and a single two-level system beyond the RWA. We evaluate the degree of squeezing of the field mode present in the ground state of the dispersive Hamiltonian and compare it with that of the ground state of the full Rabi model.  In Sec. \ref{sec:grndstsqzngprot} we present a protocol that can significantly enhance the degree of squeezing of the field mode. The influence of noise and imperfections is examined in Sec.~\ref{sec:noiseevl} and the paper concludes with a short discussion in Sec.~\ref{sec:outlk}.

\section{Dispersive regime: beyond the RWA} \label{sec:disp}

The physical setting we consider is a strongly coupled light-matter system modeled by the Rabi Hamiltonian of Eq.~\eqref{fullrabi}. Throughout this paper we focus on the dispersive regime in which the detuning $\Delta \equiv \Omega - \omega$  between the qubit and the cavity is large compared to their coupling, $\lvert \Delta \rvert \gg g$. Furthermore, we will assume that $\lvert \Delta \rvert \sim \Omega,\omega$ and therefore will not invoke the RWA to simplify the light-matter interaction. Although the theory presented here may describe many different types of experimental systems, the dispersive limit is particularly applicable to circuit QED experiments \cite{abla04,awal04,dzue09,ndidi14,melli15}, which we will touch on near the end of the paper.

\begin{figure}[t]
  \centering
    \includegraphics[width=0.49\textwidth]{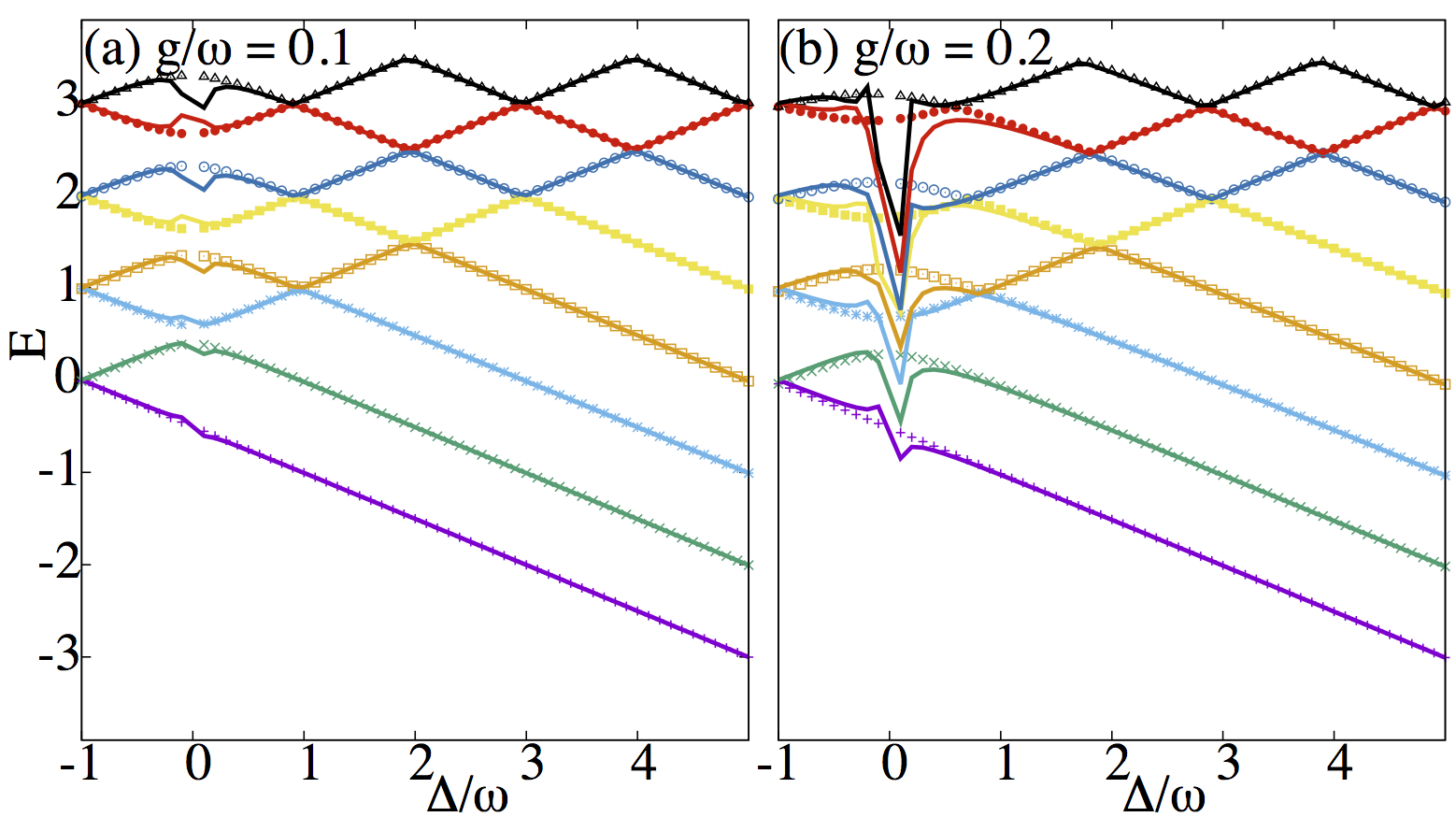}
        \caption{(Color online) Comparison between a numerical solution of the lowest energy levels of the full Rabi model [Eq.~\eqref{fullrabi}] (dotted) and the analytical solutions of the dispersive Hamiltonian [Eq.\eqref{disphm}] (solid) for (a) $g/\omega=0.1$ and (b) $g/\omega=0.2$. Note that $\Omega/\omega=1+\Delta/\omega$.}\label{eigvales}
\end{figure}

An effective Hamiltonian in the dispersive limit may be derived \cite{dzue09} using the unitary transformation $D=e^{\zeta\hat{a}^{\dagger}\sigma^{-}+\tilde{\zeta}\hat{a}^{\dagger}\sigma^{+}-h.c.}$, where $\zeta \equiv g/\Delta$, $\tilde{\zeta} \equiv g/(\Omega+\omega)$, and $h.c.$ denotes the Hermitian conjugate. Applying this transformation to the Rabi Hamiltonian and keeping terms up to second order in $g$, the dispersive Hamiltonian $H_{\rm disp}$ is given by
\begin{equation}\label{disphmold}
\begin{split}
H_{\rm disp}&=\hat{D}^{\dagger}H_{\rm Rabi}\hat{D} \\
&=\omega \hat{a}^{\dagger}\hat{a}+\frac{\Omega}{2} \hat{\sigma}_{z} +\frac{1}{2}\left(\frac{g^{2}}{\Delta}+\frac{g^{2}}{2\Omega-\Delta}\right) \hat{\sigma}_{z}(\hat{a}^{\dagger}+\hat{a})^{2}.
\end{split}
\end{equation}
The eigenspectra of the dispersive Hamiltonian \eqref{disphmold} and the full Rabi model \eqref{fullrabi} are compared in Fig.~\ref{eigvales}. The lowest lying energy levels of each model are plotted as a function of the detuning parameter $\Delta$ for two different values of the light-matter coupling strength $g$. As is clear from Fig.~\ref{eigvales}, the dispersive theory is a valid approximation to the full Rabi model in the large detuning regime but breaks down as expected near $\Delta = 0$.  The mismatch between the two becomes more pronounced for larger values of $g$ as $\Delta \rightarrow 0$. 


An alternative form of Eq.~\eqref{disphmold} makes the physics of the dispersive interaction more transparent. Defining the parameter $2\phi=g^{2}/\Delta+g^{2}/(2\Omega-\Delta)$, the dispersive Hamiltonian may be re-expressed as
\begin{equation}\label{disphm}
H_{\rm disp}=(\omega+2\phi \hat{\sigma}_{z})\hat{a}^{\dagger}\hat{a}+\left(\frac{\Omega}{2}+\phi\right) \hat{\sigma}_{z} + \phi \hat{\sigma}_{z}(\hat{a}^{2}+\hat{a}^{\dagger 2}).
\end{equation}
In this form it is evident that the first term contains a shift in the frequency of the cavity mode that depends on the state of the qubit through $\hat{\sigma}_{z}$. This is the basis for the commonly used dispersive readout technique for superconducting qubits \cite{abla04,awal04}: the frequency shift of the cavity, which is readily measured, is correlated with the qubit state. As pointed out in \cite{dzue09}, a remarkable feature of the above Hamiltonian is that a dispersive readout of the cavity field is still possible even when the cavity field and the two-level system are coupled strongly enough that the RWA cannot be made. The final term of the dispersive Hamiltonian is also of interest. It takes the form of a one-mode squeezing interaction, the sign of which again depends on the state of the qubit through $\hat{\sigma}_{z}$ \cite{dzue09}. This suggests that the dispersive Hamiltonian may be used to generate non-classical states of the field mode, which is the main focus of the present work.

To begin with, we consider the degree of squeezing intrinsic to the ground state of the dispersive Hamiltonian and, from there, the corresponding approximate ground state of the Rabi Hamiltonian \cite{jlar07,hchen89,tsand03,sashh10,sashh13}. 
Noting that $\hat{\sigma}_{z}$ is a constant of motion of the dispersive Hamiltonian, the ground state of Eq.~\eqref{disphm} can be readily obtained by diagonalizing it in the subspace of the qubit states $\lvert \uparrow \rangle, \lvert \downarrow \rangle$ which denote, respectively, the $+$ and $-$ eigenstates of $\hat{\sigma}_z$: 
\begin{equation}\label{disphmpm}
H_{\rm disp} = H_{\rm disp}^{+} \lvert \uparrow \rangle \langle \uparrow \rvert + H_{\rm disp}^{-}\lvert \downarrow \rangle \langle \downarrow \rvert,
\end{equation}
where
\begin{equation}\label{disphmpmhlf}
H_{\rm disp}^{\pm} = (\omega \pm 2 \phi) \hat{a}^{\dagger} \hat{a} \pm \left(\frac{\Omega}{2} + \phi \right) \pm \phi (\hat{a}^{2} + \hat{a}^{\dagger 2}).
\end{equation}
To diagonalize the Hamiltonians $H_{\rm disp}^{\pm}$ we define a unitary transformation
\begin{equation}
\hat{S}=\hat{S}(r_{+})\lvert\uparrow \rangle \langle \uparrow \rvert+\hat{S}(r_{-})\lvert\downarrow \rangle \langle \downarrow \rvert,
\end{equation}
where $\hat{S}(r_{\pm})=e^{(r_{\pm} \hat{a}^{\dagger 2}-r_{\pm} \hat{a}^{2})/2}$ is a unitary squeezing operator \cite{walls}. Under the action of this unitary transformation the annihilation and creation operators $\hat{a},\hat{a}^{\dagger}$ transform as 
\begin{align}
\hat{S}^{\dagger}(r_{\pm})\hat{a}\hat{S}(r_{\pm}) &= \hat{a} \cosh r_{\pm} + \hat{a}^{\dagger} \sinh r_{\pm} ,\\ 
\hat{S}^{\dagger}(r_{\pm})\hat{a}^{\dagger}\hat{S}(r_{\pm}) &= \hat{a}^{\dagger} \cosh r_{\pm} + \hat{a} \sinh r_{\pm},
\end{align}  
where the squeezing parameter is defined as
\begin{equation}\label{sqzngparm}
\begin{split}
r_{\pm}&=-\frac{1}{2} \tanh^{-1} \left( \frac{\pm 2\phi}{\omega\pm 2 \phi} \right) \\
&=\frac{1}{4} \ln \left(\frac{\omega}{\omega\pm 4\phi}\right).
\end{split}
\end{equation}
After the transformation $\hat{S}$ is applied, the Hamiltonian of Eq.~\eqref{disphmpm} takes the diagonal form
\begin{equation}\label{disphmdiag}
\tilde{H}_{\rm disp}=\tilde{H}_{\rm disp}^{+}\lvert\uparrow \rangle \langle \uparrow \rvert+\tilde{H}_{\rm disp}^{-}\lvert\downarrow \rangle \langle \downarrow \rvert,
\end{equation}
where   
\begin{equation}\label{digdisp}
\begin{split}
\tilde{H}_{\rm disp}^{\pm}&=\hat{S}^{\dagger}H_{\rm disp}^{\pm}\hat{S} \\
&=\sqrt{\omega(\omega \pm 4 \phi)}\hat{a}^{\dagger}\hat{a} \pm \left( \frac{\Omega}{2} + \phi \right). 
\end{split}
\end{equation}
The resulting Hamiltonian is that of a harmonic oscillator whose frequency shift is correlated with the qubit state $\ket{\uparrow}$ or $\ket{\downarrow}$. Clearly, in order for the cavity mode to maintain its harmonic behavior it is necessary to have $\omega \geq \lvert 4\phi \rvert$ or, equivalently, $g \leq \sqrt{\omega \lvert \Omega^2 - \omega^2 \rvert / 4 \Omega}$. This places a further restriction on the magnitude of the coupling strength $g$, which is in addition to the requirement that $g \ll \lvert \Delta \rvert$ in order for the truncation at $\mathcal{O}(g^2)$ used in deriving Eq.~\eqref{disphmold} to be valid. 

The eigenstates of $\tilde{H}_{\rm disp}$ are simply given by $\lvert \tilde{\Psi}^{n,+}_{\rm disp} \rangle = \lvert n \rangle \lvert \uparrow \rangle$ and $\lvert \tilde{\Psi}^{n,-}_{\rm disp} \rangle=\lvert n \rangle \lvert \downarrow \rangle (n=0,1,2,...)$. The dispersive Hamiltonian \eqref{disphmpm} consequently has the corresponding eigenstates     
\begin{equation}
\lvert \Psi^{n,\pm}_{\rm disp} \rangle = \hat{S} \lvert \tilde{\Psi}^{n,\pm}_{\rm disp} \rangle.
\end{equation}
Hence the ground states of the dispersive Hamiltonians $H_{\rm disp}^{\pm}$ take a separable form with the cavity field in a squeezed vacuum state.

Since a unitary transformation leaves eigenvalues unchanged, the eigenvalues of the dispersive Hamiltonian may be taken as a direct approximation to the eigenvalues of the full Rabi model. However, the same is \textit{not} true of the eigenstates \cite{lcggo16}. The eigenstates of the original Rabi model \eqref{fullrabi} are related to the eigenstates of the dispersive Hamiltonian \eqref{disphmpm} through the unitary transformations $\hat{S}$ and $\hat{D}$: 
\begin{equation}
\lvert \Psi^{n,\pm}_{\rm Rabi} \rangle \simeq \hat{D} \lvert \Psi^{n,\pm}_{\rm disp} \rangle = \hat{D} \hat{S} \lvert \tilde{\Psi}^{n,\pm}_{\rm disp} \rangle .
\end{equation} 
To maintain consistency with the dispersive Hamiltonian \eqref{disphmpm}, which is valid to second order in $g$, we expand the operator $\hat{D}$ to first order in $\zeta, \tilde{\zeta}$ (which are proportional to $g$):
\begin{equation}
\lvert\Psi^{n,\pm}_{\rm Rabi} \rangle \simeq (1 + \zeta \hat{a}^{\dagger} \sigma^{-} + \tilde{\zeta} \hat{a}^{\dagger} \sigma^{+} - h.c.) \hat{S} \lvert \tilde{\Psi}^{n,\pm}_{\rm disp} \rangle.
\end{equation}
For $\lvert \tilde{\Psi}^{0,\mp}_{\rm disp} \rangle = \lvert 0 \rangle \ket{\downarrow,\uparrow }$, the corresponding approximate eigenstates of the Rabi model are 
\begin{align}
\begin{split}\label{rabifrmneg}
\lvert \Psi^{0,-}_{\rm Rabi} \rangle &= \hat{S}(r_{-})\lvert 0 \rangle \lvert \downarrow \rangle \\
& \quad +\hat{S}(r_{-}) (\tilde{\zeta} {\rm cosh} r_{-} - \zeta {\rm sinh} r_{-}) \lvert 1 \rangle \lvert \uparrow \rangle
\end{split}
\\
\begin{split}\label{rabifrmpos}
\lvert \Psi^{0,+}_{\rm Rabi} \rangle &= \hat{S}(r_{+}) \lvert 0 \rangle \lvert \uparrow \rangle , \\
& \quad +\hat{S}(r_{+}) (\zeta {\rm cosh} r_{+} - \tilde{\zeta} {\rm sinh} r_{+})\lvert 1 \rangle \lvert \downarrow \rangle .
\end{split} 
\end{align} 
For $\Delta > 0$ the state $\ket{\Psi^{0,-}_{\rm Rabi}}$ is an approximation to the ground state of the Rabi Hamiltonian; we have numerically confirmed this for the parameter values used throughout the work. The state $\lvert \Psi^{0,+}_{\rm Rabi} \rangle$ is a little more intriguing as it corresponds to one of the higher excited states of the Rabi model; which excited state depends on the value of $\Delta$. We have numerically confirmed that for values of $g/\omega \leq 0.2$, $\lvert \Psi^{0,+}_{\rm Rabi} \rangle$ is an approximation to the $n^{\rm th}$ excited state when $n=4$ ($\Delta=2$), $n=7$ ($\Delta=5$) and $n=12$ ($\Delta=10$). 

These states are not separable as the dispersive eigenstates are; rather, they represent entangled states of the qubit and the cavity field. What is more, only the first term in each superposition, albeit the dominant one, contains a squeezed vacuum in the field. The second term is a squeezed number or Fock state, which is an even more highly nonclassical state than the squeezed vacuum. Nevertheless, a squeezed vacuum state of the field may be recovered by making a projective measurement onto the appropriate qubit state. The differences between the dispersive eigenstates and the approximate Rabi eigenstates have important consequences for the generation of squeezed states of the field, as we shall see later on.

In order to quantify the degree of squeezing of the cavity field present in the the dispersive and Rabi ground states, we introduce dimensionless position and momentum quadratures for the mode $\hat{a}$:
\begin{equation}
\begin{split}
\hat{X}_{\hat{a}} &= (\hat{a}^{\dagger} + \hat{a}) \\
\hat{P}_{\hat{a}} &= i(\hat{a}^{\dagger} - \hat{a}).
\end{split}\label{barcavopert}
\end{equation}
First we consider the ground state in each qubit subspace of the dispersive Hamiltonian, $\hat{S} \lvert 0 \rangle \lvert \uparrow, \downarrow \rangle$, for which the variances in the position and momentum quadratures are 
\begin{align}
\langle \Delta \hat{X}_{\hat{a}}^{2} \rangle &= \sqrt{\frac{\omega}{\omega \pm 4\phi}} \\
\langle \Delta \hat{P}_{\hat{a}}^{2} \rangle &= \sqrt{\frac{\omega\pm 4\phi}{\omega}}.
\end{align}
Alternatively, in terms of the squeezing parameter given by Eq.~\eqref{sqzngparm} the variances may be written as 
\begin{align}
\langle \Delta \hat{X}_{\hat{a}}^{2} \rangle &= e^{2r_{\pm}} \\
\langle \Delta \hat{P}_{\hat{a}}^{2} \rangle &= e^{-2r_{\pm}}.
\end{align} 
It is clear that, for $\phi > 0$, i.e. $\Delta > 0$, the squeezing parameter $r_{-}$ (which corresponds to the ground state of Eq.~\eqref{disphmdiag}) results in noise reduction beyond the standard quantum limit in the momentum quadrature and enhanced fluctuations in the position quadrature. The reverse is true for squeezing parameter $r_{+}$. Hence the quadrature of squeezing depends on the sign of $\langle \hat{\sigma}_z \rangle$.

\begin{figure}[h!]
  \centering
    \includegraphics[width=0.49\textwidth]{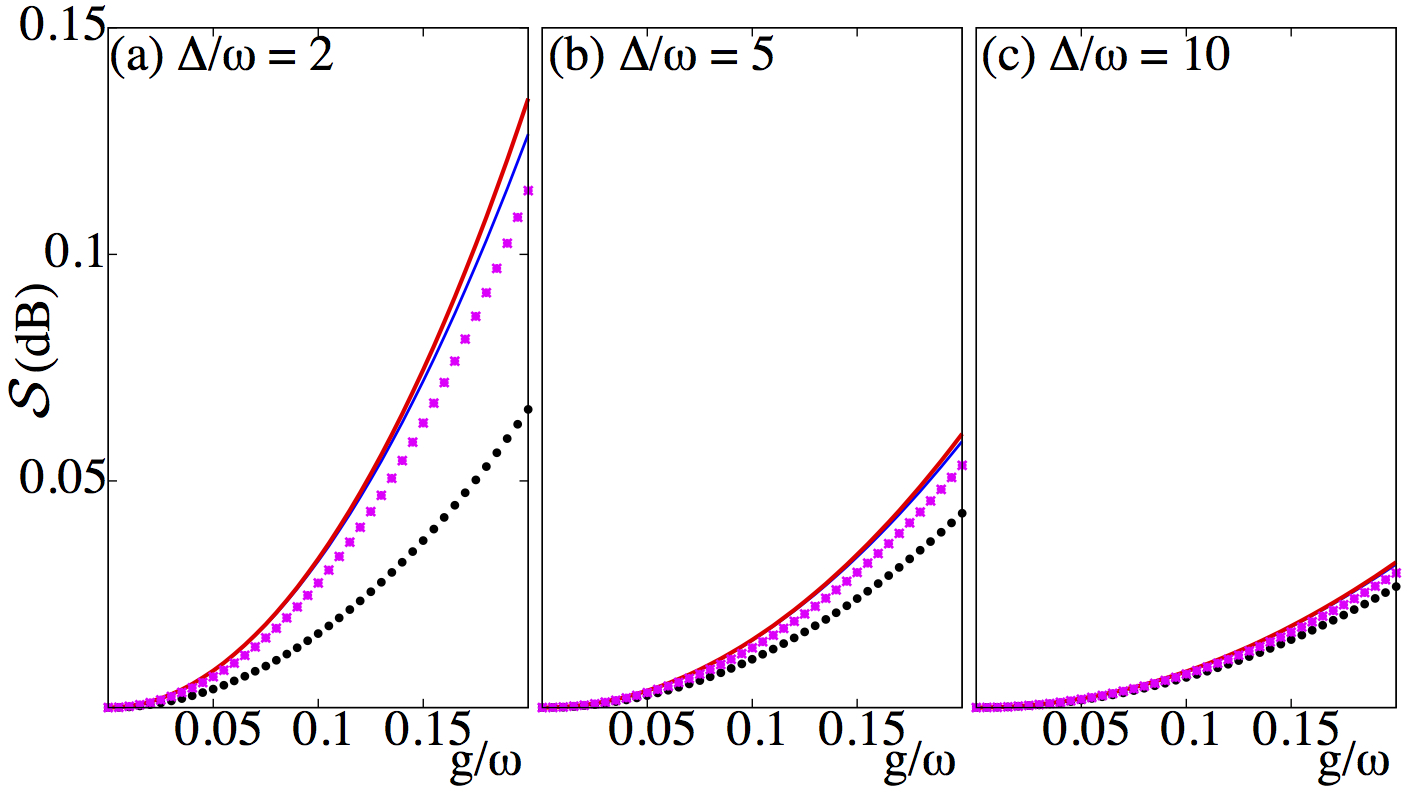}
        \caption{(Color online) Degree of squeezing $\mathcal S$ as a function of the light-matter coupling strength $g$ for the dispersive eigenstates $|\Psi^{0,+}_{\rm disp} \rangle$ (thin solid, blue)and $|\Psi^{0,-}_{\rm disp} \rangle$ (thick solid, red), the approximate ground state of the Rabi model $|\Psi^{0,-}_{\rm Rabi} \rangle$ (crosses, pink), and the numerically determined ground state of the Rabi model $|\Psi^{0}_{\rm Rabi} \rangle$ (filled-dots, black) for $(a)~\Delta/\omega=2,~(b)~\Delta/\omega=5$, and $(c)~\Delta/\omega=10$.}\label{sqznggrnd}
\end{figure}

In experiments the degree of squeezing $\mathcal{S}$ is commonly expressed in decibels (dB), calculated as $\mathcal{S} = \max(0,-10\log_{10}(\min(\langle \Delta \hat{X}_{\hat{a}}^{2} \rangle, \langle \Delta \hat{P}_{\hat{a}}^{2} \rangle)))$. The degree of squeezing present in the ground states of both qubit subspaces of the dispersive Hamiltonian $\hat{S}\ket{0} \ket{\downarrow,\uparrow}$ and in the  approximate ground state of the Rabi Hamiltonian $\ket{\Psi_{\rm{Rabi}}^{0,-}}$ is plotted in Fig.~\ref{sqznggrnd} as a function of the light-matter coupling $g$ for three different values of the detuning $\Delta$. Here and in what follows we concentrate on the case $\Delta > 0$, which gives both a larger degree of squeezing and a larger frequency shift of the cavity field for a given value of $\lvert \Delta \rvert$. Figure~\ref{sqznggrnd} shows that the degree of squeezing in the ground state of the dispersive Hamiltonian increases with the coupling $g$. This is expected since the parameter $\phi$ which controls the degree of squeezing scales as $g^2$. Similarly, the squeezing is reduced as the detuning $\Delta$ increases. Hence there is a tradeoff between the validity of the dispersive approximation, which requires $\lvert \Delta \rvert \ll g$, and the amount of squeezing that is present in the ground state. In any event, the degree of squeezing of the ground state of the dispersive Hamiltonian \eqref{disphmpm} is not large, $\mathcal{S}~<~0.1$~dB.  It can be easily verified that the squeezing parameter $\lvert r_{-} \rvert \geq \lvert r_{+} \rvert$, which is why the degree of squeezing of $\hat{S} \lvert 0\rangle \lvert \downarrow \rangle$ is marginally higher than the degree of squeezing  of $\hat{S} \lvert 0\rangle \lvert \uparrow \rangle$. 

For the sake of comparison the degree of squeezing in the exact ground state of the full Rabi model \eqref{fullrabi}, calculated numerically, is also shown in Fig.~\ref{sqznggrnd}. From the figure it is evident that as the parameter $\phi$ decreases both the dispersive ground state $\hat{S} \ket{0} \ket{\downarrow}$ and the approximate Rabi ground state $\ket{\Psi^{0,-}_{\rm Rabi}}$ better approximate the exact ground state of the Rabi model in terms of capturing the degree of squeezing $\mathcal{S}$.  The degree of squeezing in the approximate and the exact ground states of the Rabi model is smaller than the degree of squeezing present in the dispersive states $\hat{S} \ket{0} \ket{\downarrow,\uparrow}$. This can be understood by examining the form of Eq.~\eqref{rabifrmneg}, which is a superposition of a squeezed vacuum state $\hat{S}(r_-)\ket{0}\ket{\downarrow}$ and a squeezed $n=1$ Fock state $\hat{S}(r_-)\ket{1}\ket{\uparrow}$. For $n > 0$, the Fock states $\ket{n}$ are not states of minimum uncertainty; their variances are given by $_n\av{\Delta \hat{X}_{\hat{a}}^2}_n = _n\av{\Delta\hat{P}_{\hat{a}}^2}_n = 2n + 1$. Therefore the state $\hat{S}(r_-)\ket{1}$ has variances $\av{\Delta\hat{X}_{\hat{a}}^2} = 3 e^{2r_{-}}$ and $\av{\Delta\hat{P}_{\hat{a}}^2} = 3 e^{-2r_{-}}$. Since the $n=0$ and $n=1$ states are associated with orthogonal qubit components, there is no coherence between them and the overall variances are equal to the sums of the variances of the two squeezed Fock states. As a consequence, both quadrature variances are increased over those of the squeezed vacuum and hence the degree of squeezing of the overall state is reduced.

\section{Squeezing generation through sudden qubit flips}\label{sec:grndstsqzngprot}

In the previous section we evaluated the degree of squeezing in the respective ground states of the dispersive Hamiltonian and the full Rabi model. Unfortunately, the degree of squeezing $\mathcal{S}$ is relatively low even for the largest values of the light-matter coupling strength $g$ for which the dispersive approximation holds. In this section we outline a strategy to significantly improve the amount of squeezing of the oscillator mode. The basis of our protocol is a scheme detailed in Ref.~\cite{jjan92}, which uses sudden changes in the frequency of a harmonic oscillator to generate arbitrarily strong squeezing of the oscillator state. We show that this scheme can be directly realised within the dispersive Hamiltonian, using the interaction of the qubit with the field mode to create the required frequency shifts. The use of a dispersively coupled qubit distinguishes our proposal from that of Ref.~\cite{amzag08}, in which the repeated frequency shift protocol of Ref.~\cite{jjan92} was shown to produce squeezed states in a nonlinear superconducting oscillator.

To begin with we re-express the Hamiltonian \eqref{disphmpm} in terms of position and momentum coordinates for mode~$\hat{a}$:
\begin{align}
\hat{a} &= \frac{1}{2} \left(\sqrt{2\omega} \hat{x} + i\sqrt{\frac{2}{\omega}} \hat{p} \right),\\
\hat{a}^{\dagger} &= \frac{1}{2} \left(\sqrt{2\omega} \hat{x} - i\sqrt{\frac{2}{\omega}} \hat{p} \right),
\end{align}
so that the dispersive Hamiltonian of Eq.~\eqref{disphmpm} can be written as 
\begin{align}\label{xprep}
\begin{split}
H_{\rm disp} &= \left(\frac{\hat{p}^{2}}{2} + \frac{1}{2}\omega_{+}^{2} \hat{x}^{2}\right) \ket{\uparrow} \bra{\uparrow} + \left(\frac{\hat{p}^{2}}{2}+\frac{1}{2}\omega_{-}^{2} \hat{x}^{2} \right) \ket{\downarrow} \bra{\downarrow} \\
&= \omega_+ \hat{a}_+^{\dagger} \hat{a}_+ \ket{\uparrow} \bra{\uparrow} + \omega_- \hat{a}_-^{\dagger} \hat{a}_- \ket{\downarrow} \bra{\downarrow}.
\end{split}
\end{align}
Within each qubit subspace the Hamiltonian of the field takes the form of a harmonic oscillator of unit mass and shifted oscillation frequency $\omega_{\pm}^{2} = \omega^2 \pm 4 \omega \phi$. The lowering and raising operators associated with the frequency-shifted potentials are defined as
\begin{align}
\hat{a}_{\pm} &= \frac{1}{2} \left( \sqrt{2 \omega_{\pm}} \hat{x} + i \sqrt{\frac{2}{\omega_{\pm}}} \hat{p} \right), \\
\hat{a}_{\pm}^{\dagger} &= \frac{1}{2} \left( \sqrt{2 \omega_{\pm}} \hat{x} - i \sqrt{\frac{2}{\omega_{\pm}}} \hat{p} \right).
\end{align}
Note that the modes defined by $\hat{a}_{\pm}$ are not independent: both sets of operators depend on the same underlying position and momentum coordinates, so that $[\hat{a}_{+}, \hat{a}_{-}] \neq 0$. Rather, the frequency-shifted mode operators are related to the original oscillator mode by $\hat{a}_{\pm} = \hat{S}(r_{\pm}) \hat{a} \hat{S}^{\dagger}(r_{\pm})$.

Now we imagine a scenario in which the system is initialized in the ground state of the dispersive Hamiltonian $\ket{\Psi_{\rm disp}^{0,-}} = \hat{S} \ket{0} \ket{\downarrow}$. At time $t=0$ the state of the qubit $\ket{\psi_q}$ is flipped suddenly:  
\begin{align}
\ket{\psi_{q}(t)} &= \ket{\downarrow}, & t &< 0 \\
\ket{\psi_{q}(t)} &= \ket{\uparrow}, & t &\geq 0.
\end{align}
The orthogonality of the qubit states $\ket{\downarrow}$ and $\ket{\uparrow}$ implies that prior to $t=0$ the effective Hamiltonian of the field mode is $\omega_- \hat{a}^{\dagger}_{-} \hat{a}_{-}$, whereas for $t \geq 0$ the mode evolves under the effective Hamiltonian $\omega_+ \hat{a}^{\dagger}_{+} \hat{a}_{+}$. In other words, since the qubit state determines the frequency shift of the mode, flipping the qubit results in a sudden change in the frequency of the harmonic oscillator. 

\begin{figure}[h!]
  \centering
    \includegraphics[width=0.49\textwidth]{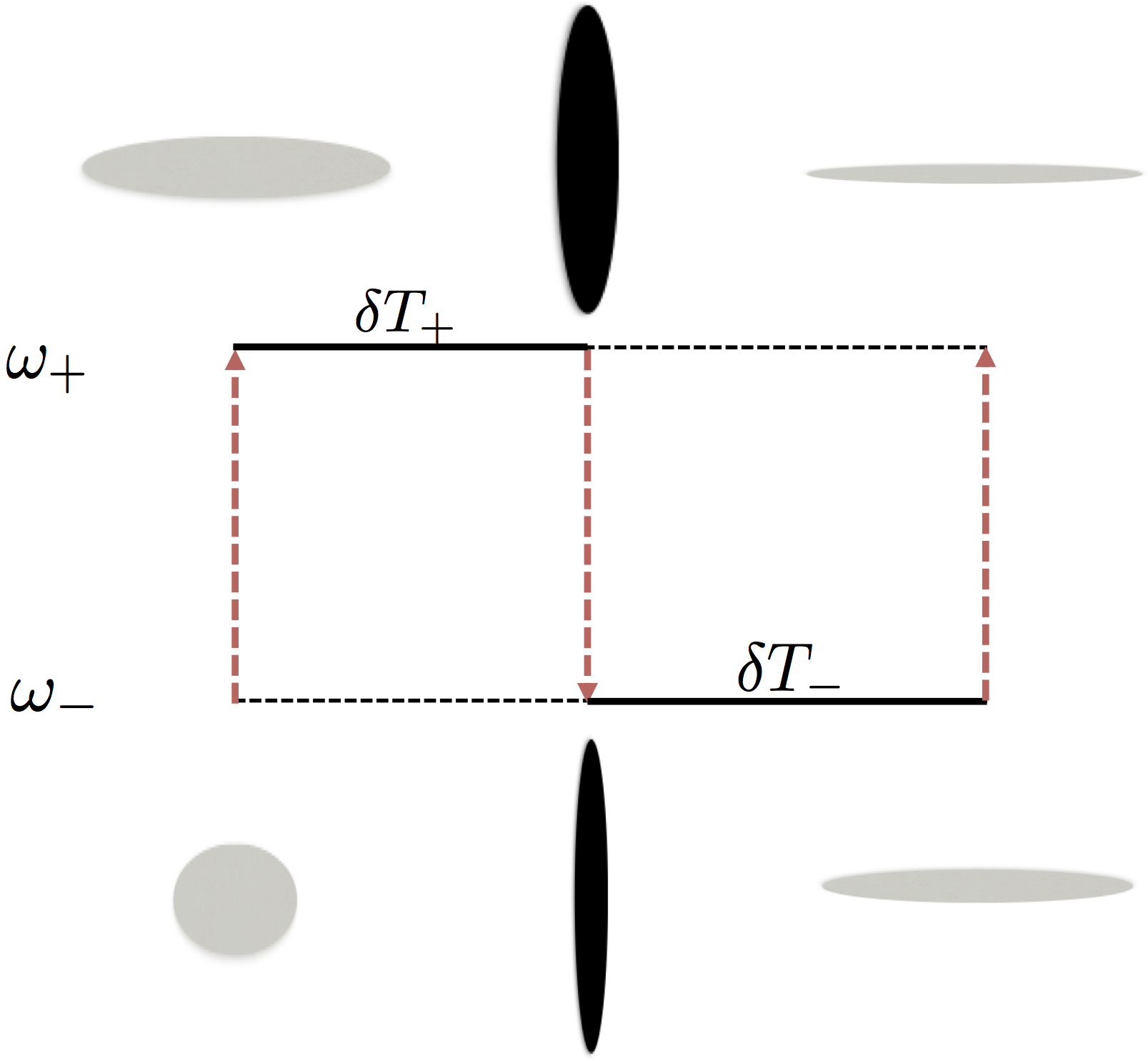}
        \caption{(Color online) Schematic of the dispersive squeezing protocol outlined in the text. Heavy dashed arrows represent qubit flips, while solid lines indicate time evolution under the corresponding oscillator Hamiltonian. The ellipses illustrate the state of the field before and after each qubit flip. One full cycle plus the first step of a second is shown here.
}\label{sqzng1cycle}
\end{figure}

It has been shown that a suitably timed sequence of sudden frequency changes is capable of generating arbitrarily large squeezing of a field mode~\cite{jjan92}. In our coupled light-matter system, the dispersive interaction together with qubit flips provides the mechanism for changing the frequency of the field mode. For the sake of completeness we briefly summarize the main steps of the protocol developed in Ref.~\cite{jjan92} within the context of our system.
\begin{itemize}
\item We assume that for $t<0$ the system is prepared in the ground state of the dispersive Hamiltonian $\ket{\Psi_{\rm disp}^{0,-}} = \hat{S} \ket{0} \ket{\downarrow} = \ket{0_-} \ket{\downarrow}$, where $\ket{0_-}$ is defined as the ground state of the frequency-shifted oscillator potential $\omega_- \hat{a}_-^{\dagger} \hat{a}_-$.
\item At $t=0$ the qubit is suddenly flipped to its excited state $\ket{\uparrow}$. Immediately following the qubit flip, although the expectation values of $\hat{x}$ and $\hat{p}$ for the field remain unchanged, the field state is squeezed \textit{relative to the new potential} $\omega_+ \hat{a}_+^{\dagger} \hat{a}_+$. 
\item The joint state evolves under the Hamiltonian $\omega_+ \hat{a}_+^{\dagger} \hat{a}_+ \ket{\uparrow}\bra{\uparrow}$ for a time duration $\delta T_{+}$. 
\item After the time delay $\delta T_{+}$, the qubit is suddenly flipped back to its ground state $\ket{\downarrow}$. This creates a second sudden frequency jump from $\omega_{+} \rightarrow \omega_{-}$, following which the state of the field is squeezed with respect to the potential $\omega_- \hat{a}_-^{\dagger} \hat{a}_-$. 
\item The first cycle of the protocol finishes with allowing the joint state to evolve under the Hamiltonian $\omega_- \hat{a}_-^{\dagger} \hat{a}_- \ket{\downarrow}\bra{\downarrow}$ for a duration $\delta T_{-}$.
\item By carefully choosing $\delta T_{\pm}$ and repeating the above steps $N$ times very strong squeezing of the field mode $\hat{a}$ can be generated. 
\end{itemize}
These steps are illustrated in Fig.~\ref{sqzng1cycle}.
As long as we are working within the dispersive Hamiltonian, the above protocol is directly analogous to the harmonic oscillator with time-dependent frequency envisioned by Janszky and Adam \cite{jjan92}. Following their Heisenberg-picture analysis, we arrive at the following expressions for the time-evolved operators $\hat{a}_{-}(\delta T_{-},\delta T_{+}),\hat{a}_{-}^{\dagger}(\delta T_{-},\delta T_{+})$ after one cycle of the protocol:
\begin{align}
\hat{a}_{-}(\delta T_{-},\delta T_{+}) &= \hat{a}_{-}(\delta T_{+}) e^{-i \omega_{-}\delta T_{-}},\\
\hat{a}_{-}^{\dagger}(\delta T_{-},\delta T_{+}) &= \hat{a}_{-}^{\dagger}(\delta T_{+}) e^{i \omega_{-}\delta T_{-}}.
\end{align}
The operators for the $\omega_-$ potential immediately following the second qubit flip are given by
\begin{align}
\begin{split}
\hat{a}_{-}(\delta T_{+}) &= \cos (\omega_{+}\delta T_{+})\hat{a}_{-}(-0) \\
& \quad -i \sin (\omega_{+}\delta T_{+})[u_{+}^{2}\hat{a}_{-}(-0)+u_{-}^{2} \hat{a}_{-}^{\dagger}(-0)] 
\end{split} \\
\begin{split}
\hat{a}_{-}^{\dagger}(\delta T_{+}) &= \cos (\omega_{+}\delta T_{+})\hat{a}_{-}^{\dagger}(-0) \\
& \quad +i \sin (\omega_{+}\delta T_{+})[u_{+}^{2}\hat{a}_{-}^{\dagger}(-0)+u_{-}^{2} \hat{a}_{-}(-0)],
\end{split}
\end{align}
where $\hat{a}_{-}^{\dagger}(-0),\hat{a}_{-}(-0)$ are the initial creation and destruction operators before the start of the protocol and $u_{\pm}=(\omega_{+}^{2}\pm\omega_{-}^{2})/(2\omega_{+}\omega_{-})$. The time evolved operators $\hat{a}_{-}(\delta T_{+}),\hat{a}_{-}^{\dagger}(\delta T_{+})$ depend crucially on the choice of $\delta T_{+}$. If $\delta T_{+}=m\pi/\omega_{+}~(m \in \mathbb{Z})$ no squeezing is generated by the protocol. Maximal squeezing is obtained when $\delta T_{+}=(2m+1)\pi/2\omega_{+}$. Likewise, if the protocol is to be further repeated it is critical to choose $\delta T_{-} = (2m+1)\pi/2\omega_{-}$. From now on we therefore fix $\delta T_{\pm}=\pi/2\omega_{\pm}$. The time evolved position and momentum quadratures defined with respect to the mode with frequency $\omega_-$ then become 
\begin{align}
\hat{X}_{-}(\delta T_{-},\delta T_{+}) &= (\hat{a}_{-}^{\dagger}(\delta T_{-},\delta T_{+})+\hat{a}_{-}(\delta T_{-},\delta T_{+})) ,\\
\hat{P}_{-}(\delta T_{-},\delta T_{+}) &= i(\hat{a}_{-}^{\dagger}(\delta T_{-},\delta T_{+})-\hat{a}_{-}(\delta T_{-},\delta T_{+})) .
\end{align}
The corresponding variances of the quadrature operators are given by
\begin{align}
\langle \Delta X_{-}(\delta T_{-},\delta T_{+})^{2}\rangle&=\left(\frac{\omega_{+}}{\omega_{-}}\right)^{2} ,\\
\langle \Delta P_{-}(\delta T_{-},\delta T_{+})^{2}\rangle&=\left(\frac{\omega_{-}}{\omega_{+}}\right)^{2},
\end{align}
which clearly illustrates that the protocol decreases quantum noise beyond the standard quantum limit in the momentum quadrature at the expense of increased fluctuations in the position quadrature of the field mode. This protocol can be repeated to produce even greater squeezing: after $N$ cycles the variances are given by $\langle \Delta \hat{X}_{-}(\delta T_{-},\delta T_{+})^{2}\rangle^N$ and $\langle \Delta \hat{P}_{-}(\delta T_{-},\delta T_{+})^{2}\rangle^N$.

Unlike the proposal in Ref.~\cite{jjan92}, the time-dependent frequency changes in our system are achieved via an effective interaction with a qubit. It makes sense, therefore, to consider the squeezing relative to the bare cavity mode with frequency $\omega$ and lowering and raising operators $\hat{a}, \hat{a}^{\dagger}$. Following one cycle of the protocol, the time-evolved quadrature operators for the bare cavity mode $\hat{X}_{\hat{a}}(\delta T_{-},\delta T_{+}),\hat{P}_{\hat{a}}(\delta T_{-},\delta T_{+})$ are related to the operators for the $\hat{a}_-$ mode by  
\begin{align}
\hat{X}_{\hat{a}}(\delta T_{-},\delta T_{+})&=e^{r_{-}}\hat{X}_{-}(\delta T_{-},\delta T_{+}) ,\\
\hat{P}_{\hat{a}}(\delta T_{-},\delta T_{+})&=e^{-r_{-}}\hat{P}_{-}(\delta T_{-},\delta T_{+}).
\end{align} 
After $N$ cycles of the protocol the variances in the position and momentum quadratures of the bare cavity mode are given by $e^{2r_{-}}\langle \Delta \hat{X}_{-}(\delta T_{-},\delta T_{+})^{2}\rangle^N$ and $e^{-2r_{-}}\langle \Delta \hat{P}_{-}(\delta T_{-},\delta T_{+})^{2}\rangle^N$, respectively. In the calculations that follow the degree of squeezing $\mathcal{S}$ is always computed with respect to the bare cavity mode.

\begin{figure}[h!]
  \centering
    \includegraphics[width=0.49\textwidth]{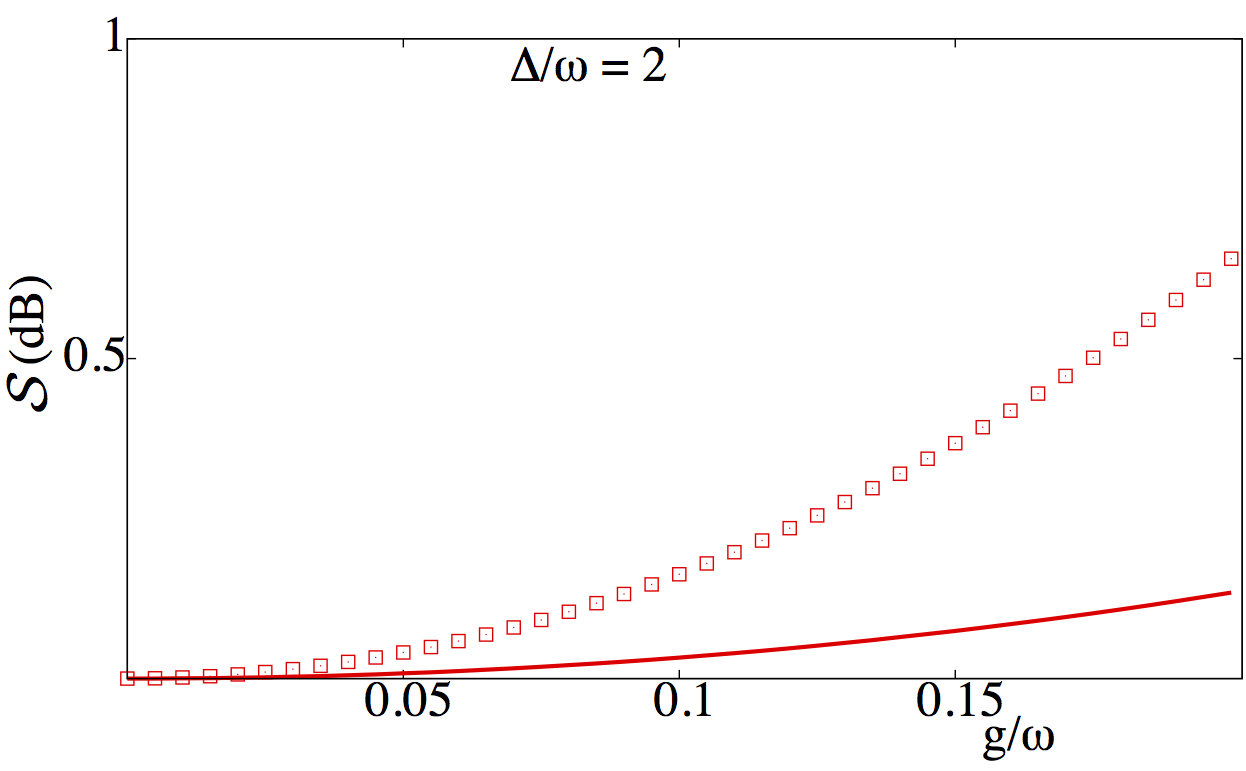}
        \caption{(Color online) Degree of squeezing obtained after one cycle of the protocol in the ideal dispersive case (open squares), as a function of the coupling strength $g/\omega$. For comparison the squeezing present in the ground state $\ket{\Psi^{0,-}_{\rm disp}}$ is also shown (solid). Other physical parameters include: $\Delta/\omega=2, \Omega/\omega=1+\Delta/\omega$.}\label{compdispjumpprot}
\end{figure}

Figure~\ref{compdispjumpprot} compares the degree of squeezing present in the ground state $\ket{\Psi^{0,-}_{\rm disp}}$ with that obtained after one cycle of the protocol, as a function of the coupling strength $g/\omega$. It can be clearly seen that the protocol using sudden frequency flips can significantly increase the degree of squeezing of the field mode over that naturally present in the ground state. 


The foregoing discussion has been based on the dispersive Hamiltonian and its eigenstates. However, in our system the dispersive Hamiltonian arises as an approximation to the full Rabi Hamiltonian. As the squeezing protocol involves manipulating the state of the system, working with the Rabi Hamiltonian itself will give different results than working with the dispersive Hamiltonian. In order to analyze the outcome of the protocol using the Rabi Hamiltonian, it is useful to work in the Schr\"{o}dinger picture rather than the Heisenberg picture used previously.

In the Schr\"{o}dinger picture, one cycle of the protocol based on the dispersive Hamiltonian as described above results in the state
\begin{equation}
\begin{split}
\ket{\Psi_{\rm disp}(\delta T_-, \delta T_+)} &= e^{-iH_{\rm disp} \delta T_{-}} \sigma_{x} e^{-iH_{\rm disp} \delta T_{+}}\sigma_{x} \hat{S}(r_{-})\ket{0} \ket{\downarrow} \\
&= e^{-iH^{-}_{\rm disp} \delta T_{-}}e^{-iH^{+}_{\rm disp} \delta T_{+}}\hat{S}(r_{-})\ket{0} \ket{\downarrow}.
\end{split}
\end{equation}
The protocol may be carried out similarly using the Rabi Hamiltonian; however, this does result in some deviation from the ideal case using the dispersive Hamiltonian. For simplicity, we assume that the initial state for the Rabi protocol is the ground state of the Rabi Hamiltonian. As is evident from the approximate solution in Eq.~\eqref{rabifrmneg}, this state is not an eigenstate of the qubit operator $\sigma_z$; rather than being a separable state of qubit and field with the qubit in $\ket{\downarrow}$, the Rabi ground state has a component along $\ket{\uparrow}$. However, provided that the parameter regime is chosen such that the dispersive approximation holds, the $\ket{\uparrow}$ component is small and will not have a significant negative impact on the degree of squeezing produced by the protocol. 

\begin{figure}[h!]
  \centering
    \includegraphics[width=0.49\textwidth]{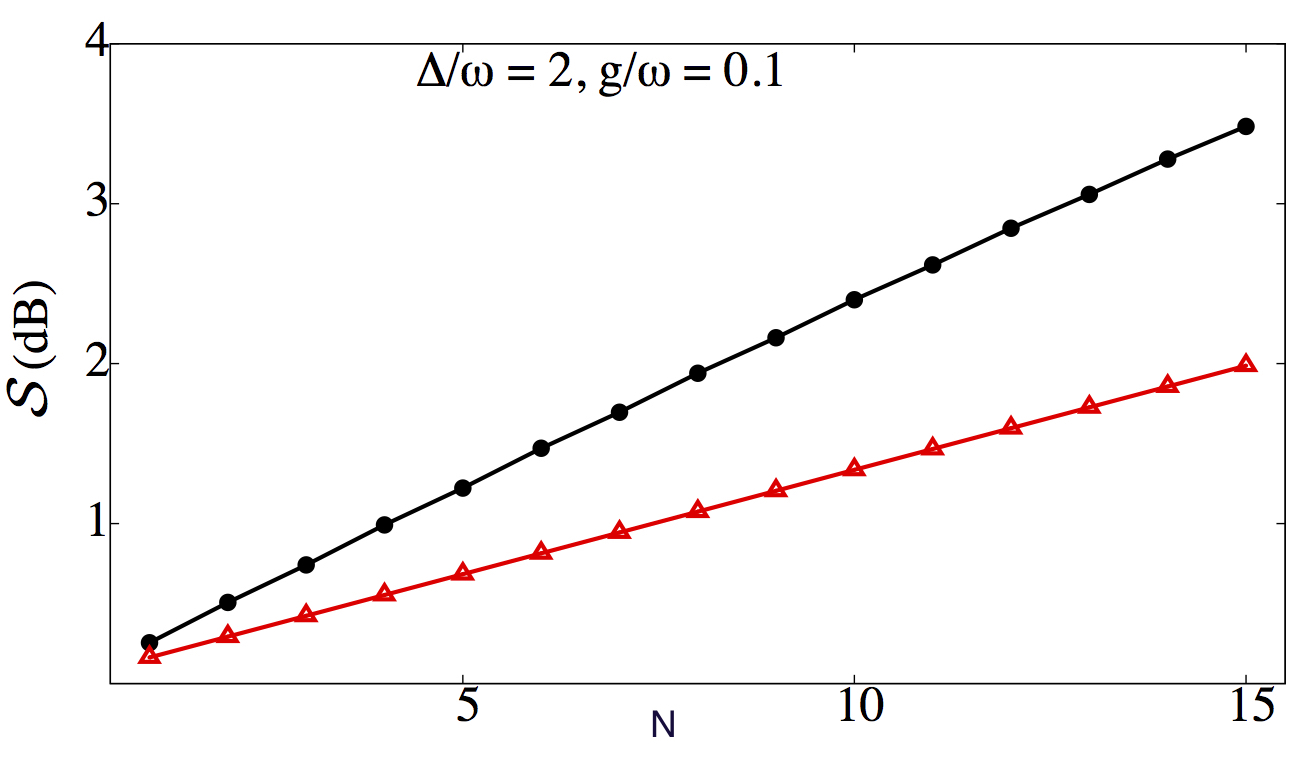}
   \caption{(Color online) Degree of squeezing $\mathcal S$ for the dispersive (red, triangles) and Rabi (black, circles) versions of the squeezing protocol, as a function of the number of cycles $N$. 
	Other physical parameters include: $g/\omega = 0.1, \Delta/\omega=2, \Omega/\omega=1+\Delta/\omega$.}\label{fiddisprabi}
\end{figure}

As in the dispersive case, the first step of the protocol is to flip the state of the qubit by applying the $\sigma_x$ operator. Of course, this does not result in a state that is purely along $\ket{\uparrow}$, but again the error induced is small. The flipped state is now allowed to evolve under the full Rabi Hamiltonian for a time $\delta T_+$. Following this evolution, the qubit is flipped again and the state evolves, still under the full Rabi Hamiltonian, for a time $\delta T_-$. The evolution time intervals $\delta T_{\pm}$ remain the same as in the dispersive case, to a good approximation. This is because the evolution time is related to the differences in energy eigenvalues, which are effectively the same in the dispersive and Rabi cases provided that the parameters are chosen suitably. The resultant state after one cycle of the protocol is then given by 
\begin{equation}\label{timevolbrabi1cycl}
\ket{\Psi_{\rm Rabi}(\delta T_-, \delta T_+)} = e^{-iH_{\rm Rabi} \delta T_{-}}\sigma_{x}e^{-iH_{\rm Rabi} \delta T_{+}}\sigma_{x} \ket{\Psi^{0}_{\rm Rabi}}.
\end{equation}
Of course, the protocol may then be repeated multiple times in a similar fashion. Figure~\ref{fiddisprabi} shows the results of numerical calculations of the degree of squeezing $\mathcal{S}$ that is produced as a function of the number of cycles $N$ for both the dispersive and Rabi cases. It is clear that in both cases the degree of squeezing increases linearly with the number of cycles. 

Remarkably, Fig.~\ref{fiddisprabi} shows that the protocol using the Rabi Hamiltonian produces considerably more squeezing per cycle (approximately twice as much, for these parameters) than the ``ideal'' dispersive case. This is particularly surprising since the ground state of the Rabi model shows less squeezing than the dispersive ground state, which can be attributed to the higher variance of the mode state associated with the $\ket{\uparrow}$ state of the qubit as discussed at the end of Sec.~\ref{sec:disp}. Calculations show that the $\ket{\downarrow}$ component of the state is squeezed much more strongly after one cycle of the Rabi protocol than predictions based on the dispersive analysis would suggest. The variance in $\hat{P}_{\hat{a}}$ of the $\ket{\uparrow}$ state is also reduced by the squeezing protocol, albeit not below the standard quantum limit of $\langle \Delta \hat{P}_{\hat{a}}^{2} \rangle = 1$. However, the overall increase in squeezing in the Rabi case can be attributed to the very large degree of squeezing of the $\ket{\downarrow}$ component. Clearly, although the dispersive theory gives good predictions for the energies and the degree of squeezing in the ground state in the Rabi model, it is not particularly good at predicting the results from the Rabi squeezing protocol. This counterintuitive result highlights the need to be cautious when applying the dispersive approximation in situations where the state of the system is being manipulated \cite{lcggo16}.

\section{Imperfections in the protocol}
\label{sec:noiseevl} 

In a realistic setting the protocol presented in the previous section will suffer from losses and imperfections. Qubit dephasing and relaxation, cavity losses, timing jitter, and the inability to instantaneously flip the state of the qubit will all affect the outcome of the protocol. We briefly discuss each of these sources of error with an eye toward circuit QED experiments, but their relative contributions will depend on the particular experimental implementation.

In the dispersive version of the protocol, the qubit is always in one of its eigenstates, meaning that the protocol is unaffected by dephasing and only the energy relaxation time $T_1$ needs to be considered. The Rabi case is a little more complicated and qubit dephasing may contribute to noise. However, it is important to note that the qubit need only remain coherent over the evolution time interval $\delta T_{\pm}$ rather than throughout the full $N$ cycles of the protocol. To take some numbers relevant to superconducting circuit QED experiments, choosing a cavity frequency $\omega \sim 1$~GHz and a coupling strength $g=0.1~\omega$ gives $\delta T_{\pm}$ on the order of nanoseconds. Given that superconducting qubits are now routinely achieving relaxation and dephasing times of several tens of microseconds \cite{Rigetti2012,Barends2013,Braumuller2016,Yan2016}, it is clear that qubit losses are not expected to be a limiting factor for the protocol or the degree of squeezing that can be achieved. 

The loss of photons from the cavity has a more severe effect. To incorporate photon losses in the squeezing protocol we use a standard master equation approach \cite{walls}, assuming that during the time intervals $\delta T_{\pm}$ the field mode couples to a zero temperature reservoir with damping rate $\Gamma$. Figure 6 shows the results of a master equation simulation with $\Gamma = 0.01/\delta T_{+}$. In the presence of photon loss the degree of squeezing is no longer linear in the number of cycles $N$; the additional squeezing generated by another round of the protocol begins to saturate as $N$ increases. However, a substantial increase in squeezing over that present in the initial state can still be obtained. It is worth noting that this value of $\Gamma$ corresponds to a cavity Q of about 150, which is two orders of magnitude less than Q-factors routinely achieved for microwave resonators in circuit QED \cite{Rigetti2012,Barends2013,Rouxinol2016} and similar to the value recently measured for a qubit-coupled nanomechanical resonator \cite{Rouxinol2016}. 

\begin{figure}[h!]
  \centering
    \includegraphics[width=0.5\textwidth]{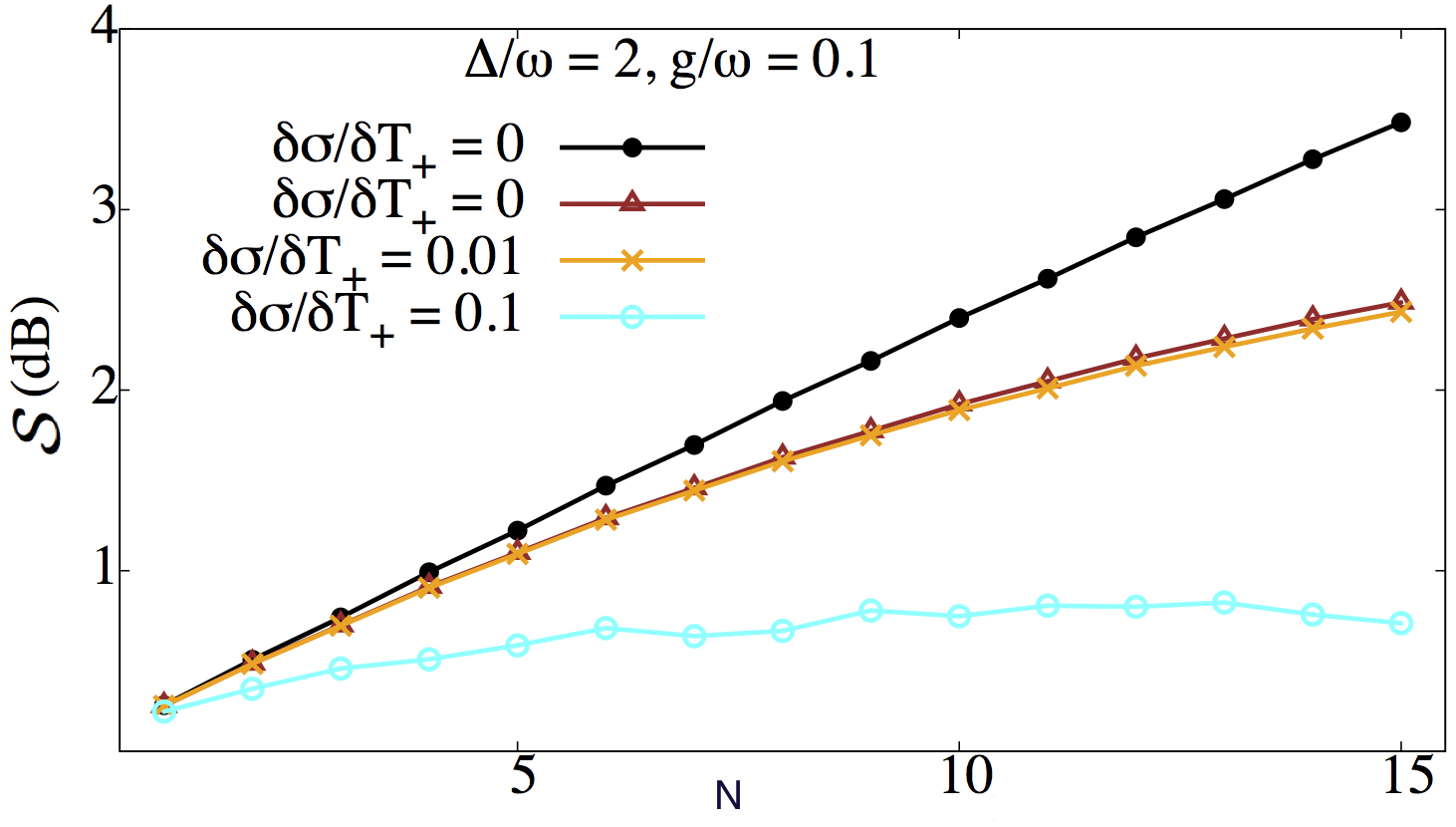}
        \caption{(Color online) Effect of photon losses and timing jitter on the average degree of squeezing $\langle \mathcal S \rangle$ produced by $N$ cycles of the Rabi protocol. The three lower curves show the effect of photon loss together with three different degrees of timing imperfections: photon loss alone with no timing error (brown, triangles), photon loss plus random timing errors chosen from a Gaussian distribution of width $\delta\sigma/\delta\Gamma_{\pm} = 0.01$ (yellow, crosses), and photon loss plus random timing errors with distribution width $\delta\sigma/\delta\Gamma_{\pm} = 0.1$ (cyan, open circles). For comparison, the ideal case is also shown (black, filled dots). The photon loss rate is $\Gamma = 0.01/\delta T_{+}$; other physical parameters are given by $\Delta/\omega=2, g/\omega = 0.1, \Omega/\omega = 1 + \Delta/\omega$.}\label{errorwithtime}
\end{figure}

The protocol also requires carefully designed time delays to ensure maximum squeezing of the field mode. In order to model the effect of inaccuracies in timing, we add a random offset $\epsilon$, chosen from a normal distribution with zero mean and standard deviation $\delta \sigma$, to each time delay $\delta T_{\pm}$. The average degree of squeezing of the field mode $\langle \mathcal S \rangle$ is evaluated by taking an ensemble average over multiple runs of the $N$-cycle protocol. Figure~\ref{errorwithtime} shows the average degree of squeezing $\langle \mathcal S \rangle$ when photon losses and random time delay are incorporated in our protocol. When the error in timing is on the order of 1~\%, the timing jitter has little effect on the degree of squeezing. However, larger timing errors (on the order of 10~\%) dramatically reduce the degree of squeezing produced. Hence the ability to control the timing of qubit flips to a reasonably precise degree will be important for experimental implementations of our protocol.

The foregoing calculations have relied on the assumption of instantaneous flips of the qubit state, which produce sudden changes in the frequency of the oscillator. A sudden frequency change is of course an idealization for a finite but small switching time $\tilde{t}$. As long as $\tilde{t} \ll {\rm min}(\omega_{+},\omega_{-})/|\omega_{+}^{2}-\omega_{-}^{2}|$, a sudden frequency change from $\omega_{+}\rightarrow \omega_{-}$ is a good approximation \cite{jjan92}. It should be pointed out that $\tilde{t} \sim 1/g$; therefore the requirement for very small values of $\tilde{t}$ can be relaxed by decreasing the light-matter coupling $g$ and increasing the number of cycles $N$ to achieve a similar degree of squeezing of the cavity field. Moreover, modifications to the sudden frequency shift scheme of Ref.~\cite{jjan92} have been discussed in the literature. The case of a sinusoidal frequency modulation was studied in Ref.~\cite{cmwil11}, and a Fourier-modified Janszky-Adam scheme for improved nonadiabatic generation of squeezed photons was proposed in Ref.~\cite{smats15}. While these schemes are not as efficient as the original sudden jump protocol, they are less demanding from an experimental standpoint and still produce significant squeezing of the cavity field.

\section{Outlook}
\label {sec:outlk}
In this work we have theoretically explored the potential for generating strong squeezing of a boson field mode interacting with a two-level system in the dispersive regime, without making the RWA. Although the ground state of this hybrid quantum system exhibits squeezing of the field, the degree of squeezing is not large. However, the dispersive frequency shift allows the frequency of the cavity mode to be changed by flipping the state of the qubit. A protocol based on a series of suitably timed sudden frequency jumps can be used to produce an arbitrarily large degree of squeezing in the absence of noise. Even in the presence of a realistic level of noise and experimental imperfections, the degree of squeezing produced by this protocol can be significant. One possible advantage of this method of squeezing generation compared to the use of a parametric amplifier and/or other nonlinearity based methods  is that under our protocol the degree of squeezing and the time of generation are directly controlled by the number of frequency shifts applied.

\acknowledgments{}
CJ is supported by a York Centre for Quantum Technologies (YCQT) Fellowship. We would like to acknowledge useful discussions with Borja Peropadre, Frank Deppe, Eran Ginossar, and Matthew Elliott.

\end{document}